\documentstyle[amssymb,12pt]{amsart}
\newtheorem{thm}{Theorem}
\newtheorem{cor}{Corollary}
\newtheorem{lem}{Lemma}
\newtheorem{prop}{Proposition}

\theoremstyle{definition}

\theoremstyle{remark}

\numberwithin{equation}{section}

\newcommand{\thmref}[1]{Theorem~\ref{#1}}
\newcommand{\secref}[1]{\S\ref{#1}}

\newcommand{\propref}[1]{Proposition~\ref{#1}}
\newcommand{\corref}[1]{Corollary~\ref{#1}}

\newcommand{\nc}{\newcommand}
\nc{\on}{\operatorname}
\nc{\ch}{\on{ch}}
\nc{\bw}{\bold{w}}
\nc{\be}{\bold{e}}
\nc{\osum}{\oplus}
\nc{\vk}{V^k}
\nc{\cvk}{\on{ch} V^k}
\nc{\su}{\widehat{\goth{sl}_2}}
\nc{\beq}{\begin{equation}}
\nc{\eeq}{\end{equation}}
\nc{\vM}{L(1,x)}
\nc{\limq}{\lim_{q\arr 1}}
\nc{\trans}{\top}
\nc{\onw}{\operatornamewithlimits}
\nc{\Z}{{\Bbb Z}}
\nc{\C}{{\Bbb C}}
\nc{\Q}{{\Bbb Q}}
\nc{\bbe}{\bold{1}}
\nc{\bmu}{\bold{u}}
\nc{\Pro}{{\Bbb P}^1}
\nc{\R}{{\Bbb R}}
\nc{\cond}{|\,}
\nc{\al}{\alpha}
\nc{\de}{\delta}
\nc{\til}{\tilde}
\nc{\bi}{\bibitem}
\nc{\ep}{\epsilon}
\nc{\la}{\lambda}
\nc{\om}{\omega}
\nc{\th}{\theta}
\nc{\lar}{\longrightarrow}
\nc{\arr}{\rightarrow}
\nc{\pik}{\frac{\pi^2}{6}}
\nc{\pa}{\partial}
\nc{\wt}{\widetilde}
\nc{\fel}{\frac{1}{2}}
\nc{\k}{h^\vee}
\nc{\cross}{\times}
\nc{\ca}{{\goth F}}
\nc{\Cone}{{\goth C}}
\nc{\bcone}{{\goth B}}
\nc{\onen}{1,\ldots,n}

\begin{document}
\title[TBA and dilogarithm identities]{Thermodynamic Bethe Ansatz and
dilogarithm identities I.}\thanks{Partially supported by NSF Grants}
\author{Edward Frenkel}
\address{Department of Mathematics, Harvard University, Cambridge, MA
02138}
\author{Andr\'as Szenes}
\address{Department of Mathematics, Massachusetts Institute of
Technology, Cambridge, MA 02139}

\maketitle

\section{Introduction}

In the decade that has passed since the seminal work \cite{BPZ} by
A.A.~Belavin, A.P.~Po\-lyakov, and A.B.~Zamolodchikov, a lot of progress
has been made in understanding Conformal Field Theories (CFTs) in two
dimensions. The success in the study of CFTs is due to their invariance
with respect to the Virasoro algebra, or, more generally, extended
conformal algebras. This property allows one to describe CFTs in terms
of representation theory of infinite-dimensional Lie algebras or vertex
operator algebras, and algebraic geometry of complex curves.

In \cite{Zam} A.B.~Zamolodchikov introduced an interesting class of 2D
quantum field theories -- perturbations of CFTs by relevant
operators. These theories lack conformal invariance, but possess some other
remarkable algebraic structures, which are yet to be fully understood from
the mathematical point of view. One of the properties is the existence of
infinitely many local integrals of motion in involution. This was
conjectured in \cite{Zam} (see also \cite{im}) and proved in
\cite{FF:toda}. Thus, a perturbation of a CFT is an integrable 2D quantum
field theory, and as such, it is governed by a purely elastic $S$--matrix,
which satisfies various algebraic constraints \cite{ZZ}. These constraints
are so strong that knowing the spins of local integrals of motion one can
often conjecture the $S$--matrix and hence determine the theory completely,
see \cite{Zam,KM} and references therein.

The Thermodynamic Bethe Ansatz (TBA) is a method of verifying these
conjectures, which was first applied in this context by Al.B.~Zamolodchikov
\cite{AlB1}. One starts with an integrable field theory conjectured to be
the perturbation of a CFT ${\cal T}$, and studies its ultraviolet (UV)
behavior. A theory on an infinitely long cylinder of circumference $R$ is
described by a system of integral equations called the TBA equations. To
write down this system explicitly, let us assume that the theory has $N$
species of particles with masses $m_a, a=1,\ldots,N$. One is interested in
the functions $\ep_a(\theta)$, which are called the spectral densities of
particles of species $a$, see e.g. \cite{KM}. These are functions of the
rapidity $\theta$ (recall that rapidity is related to the energy $E$ and
the momentum $p$ by the formulas $E = m \on{cosh} \theta, p = m \on{sinh}
\theta$). The TBA equations on the functions $\ep_a(\theta)$ read:
\begin{equation}    \label{tba}
m_a R \on{cosh} \theta = \ep_a(\theta) + \frac{1}{2\pi} \sum_{b=1}^N
\int_{-\infty}^\infty d \theta' \; \phi_{ab}(\theta-\theta') \; \log \left(
1 + Y_a(\theta') \right),
\end{equation}
where $Y_a(\theta) = e^{-\ep_a(\theta)}$, $\phi_{ab}(\theta) =
\displaystyle - i \frac{\pa \log S_{ab}}{\pa \theta}$, and $S_{ab}(\theta)$
is the $S$--matrix.

The TBA equations are usually hard to solve, but one can extract a lot of
information from them even without solving them explicitly. The ground
state energy of the theory is given by
\begin{equation}    \label{energy}
E(R) = -\frac{1}{2\pi} \sum_{a=0}^N \int_{-\infty}^\infty d \theta \; m_a R
\on{cosh} \theta \; \log \left( 1 + Y_a(\theta) \right).
\end{equation}
In the UV limit $R\arr 0$, in which one is supposed to recover the initial
CFT ${\cal T}$, we should have $E(R) \simeq -\pi \wt{c}(R)/6R$, where
$\wt{c}(R) \sim \wt{c} + O(R)$. From \eqref{energy} one finds using the TBA
equations, see e.g. \cite{KM}:
\begin{equation}    \label{di}
\frac{\pi^2}{6} \wt{c} = \sum_{a=1}^N L \left( \frac{1}{1+y_a}
\right),
\end{equation}
where $L(z)$ is the Rogers dilogarithm function \cite{Lewin}:
\begin{equation}    \label{rogers}
L(z) = \fel \int_0^z \left( \log w\,d\log (1-w) - \log (1-w)\,d w
\right) ,\quad 0\leq z\leq 1,
\end{equation}
and $y_a = \lim_{R\arr 0} Y_a(\theta)$. The numbers $y_a$ satisfy the
system of algebraic equations
\begin{equation}    \label{limiteq}
y_a = \prod_{b=1}^N \left( 1+\frac{1}{y_b} \right)^{N_{ab}},
\end{equation}
where $N_{ab}$ is the number of poles of $S_{ab}(\theta)$ in the upper half
plane; in particular, they do not depend on $\theta$.

If the conjectural description of the perturbation of the CFT ${\cal T}$ is
correct, the number $\wt{c}$ in the left hand side of the formula
\eqref{di} should coincide with the effective central charge of ${\cal
T}$. But in that case formula \eqref{di} can be considered as a dilogarithm
identity, which relates a rational number $\wt{c}$ to the algebraic numbers
$y_a$'s.

Many dilogarithm identities have been discovered this way in recent
years. While the TBA method has not yet been made rigorous, the identities
have been proved rigorously by other methods,
see \cite{Lewin,proof,FS}. Mathematically, the dilogarithm identities
manifest a connection between 2D quantum field theory on the one hand and
algebraic $K$--theory and number theory on the other, see \cite{FS}. We
hope that better understanding of the TBA will enable us to gain new
insights into this connection.

Recently, F.~Gliozzi and R.~Tateo \cite{GT} made an important step in
this direction. They found functional analogues of the identities
\eqref{di} for a large class of theories, which are labeled by pairs
$(G,H)$ of Dynkin diagrams of types $ADE$ and $T$ (the latter is the
diagram of type $A$ with a loop attached to one of the end vertices). In
such a theory, the species of particles are labeled by pairs of indices
$a=1,\ldots,r_G$, and $b=1,\ldots,r_H$, where $r_G$ and $r_H$ are the
numbers of vertices in the diagrams $G$ and $H$, respectively.

The main fact, which is due to Al.B.~Zamolodchikov \cite{AlB2} and, in
the general case, to F.~Ravanini, A.~Valleriani and R.~Tateo \cite{RVT}
is that any
solution $\{ Y_a^b(\theta) \}$ of the TBA equations \eqref{tba}
corresponding to the $(G,H)$ theory satisfies the following system of
algebraic equations:
\begin{equation}    \label{Y}
Y_a^b \left( \theta + \frac{\pi I}{\k_G} \right) Y_a^b \left( \theta -
\frac{\pi I}{\k_G} \right) = \prod_{c=1}^{r_G} \left( 1+ Y_c^b(\theta)
\right)^{G_{ac}} \prod_{d=1}^{r_H} \left( 1+ \frac{1}{Y_a^d(\theta)} \right)
^{-H_{bd}},
\end{equation}
where $I = \sqrt{-1}$, $(G_{ac})$ and $(H_{bd})$ are the adjacency matrices
of the diagrams $G$ and $H$, respectively, and $\k_G$ is the dual Coxeter
number of $G$.

The $Y$--{\em system} \eqref{Y} and closely related to it $T$--system play
an important role in quantum field theory and statistical mechanics
\cite{RVT,BR,KP,KNS}. In \cite{BLZ} it was conjectured that a certain class
of solutions of this system parametrizes the spectra of commuting integrals
of motion acting on minimal representations of conformal algebras.

Al.B.~Zamolodchikov \cite{AlB2} has conjectured an important periodicity
property of solutions of the system \eqref{Y}:
\begin{equation}    \label{periodicity}
Y_a^b \left( \theta + \pi I \frac{\k_G+\k_H}{\k_G} \right) =
Y_{\bar{a}}^{\bar{b}}(\theta),
\end{equation}
where $\k_H$ is the dual Coxeter number of $H$, and $\bar{a},\bar{b}$ are
the vertices conjugate to $a,b$, respectively. This periodicity property
allows one to find the conformal dimension of the field responsible for the
perturbation of the corresponding CFT, see \cite{AlB2}.

Now we can write down the dilogarithm identities conjectured in
\cite{GT}. Let $\{ Y_a^b(\theta) \}$ is a solution of the $Y$--system
\eqref{Y}. Fix $\theta$ and set $$X_a^b(m) = \frac{Y_a^b \left( \theta +
\pi I m/\k_G \right)}{1+Y_a^b \left( \theta + \pi I m/\k_G \right)}.$$
Suppose that all $X_a^b(m)$ are real numbers between $0$ and $1$. Then
\begin{equation}    \label{gt}
\sum_{a=1}^{r_G} \sum_{b=1}^{r_H} \sum_{m=1}^{\k_G+\k_H} L \left(
X_a^b(m) \right) = \frac{\pi^2}{6} r_G r_H
\k_G.
\end{equation}

Let $y_a^b = \lim_{\theta \arr +\infty} Y_a^b(\theta)$. In the limit
$\theta \arr +\infty$ the system \eqref{Y} becomes a system of the type
\eqref{limiteq}, and the identity \eqref{gt} becomes equivalent to the
identity \eqref{di} corresponding to the UV limit of the $(G,H)$ theory (in
order to relate them, one has to use the Euler identity $L(z) + L(1-z) =
\pi^2/6$). Therefore the identities \eqref{gt} can be viewed as
functional analogues of the known dilogarithm identities \eqref{di}. It is
interesting that the identity corresponding to the $(A_1,A_1)$ theory is
the Euler identity above, and the identity corresponding to the $(A_2,A_1)$
theory is the pentagon identity of the dilogarithm function, see \cite{GT}.

There are many indications that analogues of the $Y$--system can be defined
for other integrable field theories and that there are dilogarithm
identities associated to them, see \cite{T}.

In this paper, we give a proof of the periodicity conjecture
\eqref{periodicity} and the identities \eqref{gt} and their generalizations
for the $(A_n,A_1)$ theories. We also prove analogous identities for the
Bloch-Wigner function, which is the imaginary counterpart of the Rogers
dilogarithm. Our proof of these identities relies on a universal property
of the dilogarithm functions, which for the Bloch-Wigner function was first
proved by S.~Bloch \cite{Bloch}.

Our approach can be generalized to the identities corresponding to more
general diagrams. We have already obtained a complete proof of periodicity
and dilogarithm identities of $(A_n,A_2)$ type and partial results in the
general case. We will report on those results in the second part of this
paper.

Upon completing this work, we learned about the paper \cite{GT1}, in which
another approach to the dilogarithm identities \eqref{gt} was proposed and
a proof, different from ours, of the periodicity \eqref{periodicity} and
the identities \eqref{gt} for the $(A_n,A_1)$ theories was outlined.

The paper is organized as follows. In Sect.~2 we prove the periodicity
property of the $Y$--system. In Sect.~3 we give a general form of the
dilogarithm identities for the Rogers and the Bloch-Wigner dilogarithms. In
Sect.~4 we prove the identities \eqref{gt} of $(A_n,A_1)$ type and their
generalizations.

\section{Periodicity.}    \label{per}
In the case of $(A_n,A_1)$ theory the $Y$--system \eqref{Y} takes the
form:
\begin{equation}    \label{Y1}
Y_a \left( \theta + \frac{\pi I}{n+1} \right) Y_a \left( \theta -
\frac{\pi I}{n+1} \right) = (1+ Y_{a-1}(\theta))(1+Y_{a+1}(\theta)),
\end{equation}
where we put $Y_a(\theta)=Y_a^1(\theta)$.

In this section we prove the periodicity property of this system.

Let us fix $\theta$ and denote
$$Y(i,j) = Y_j \left( \theta + \pi I \frac{i}{n+1} \right),i
\in \Z, j=\onen.$$
In this notation the system \eqref{Y1} reads \beq
\label{ysys} Y(i-1,j)Y(i+1,j)=(1+Y(i,j-1))(1+Y(i,j+1)),
\end{equation}
$i \in \Z, j=\onen$, where we put $Y(i,0)=Y(i,n+1)=0$ for all $i$.

\smallskip
\noindent{\em Remark.}
In this paper we treat the system \eqref{ysys} as a system of algebraic
equations on the variables $Y(i,j)$. We do not restrict ourselves to the
special case when $Y(i,j)$ are values of functions $Y_j(\theta)$ satisfying
the $Y$--system \eqref{Y1} and therefore we do not use any global
properties of functional solutions $Y_j(\theta)$ of \eqref{Y1}.\qed
\smallskip

\begin{thm}    \label{peri}
Suppose that the variables $Y(i,j), i\in\Z, j=\onen$, satisfy the system of
equations \eqref{ysys}. Then $Y(i,j)=Y(i+n+3,n+1-j)$.
\end{thm}
{\it Proof.}  It will be helpful to pass to the new variables
$X(i,j)=Y(i,j)/(1+Y(i,j))$. In these, \eqref{ysys} takes the form
\beq \label{xsys} \left(1-\frac{1}{X(i-1,j)}\right)
\left(1-\frac{1}{X(i+1,j)}\right) = (1-X(i,j-1)) (1-X(i,j+1)).
\end{equation}

Also introduce the transformation $S:(i,j)\arr (i+n+3,n+1-j)$. Thus our
goal is to show that $Y(i,j)=Y(S(i,j)$).

The method of proof is setting up some initial conditions and then solving
the system of equations explicitly. Note that the variables $Y(i,j)$ and
$X(i,j)$ with $i+j$ even are independent from those with $i+j$
odd. Therefore without loss of generality we can restrict ourselves to
those $X(i,j)$ for which $i+j$ is even.

Introduce a set of variables $a_1, a_2,\dots,a_n$, and set $X(i,i)=a_i$ for
$i=1,\dots n$. The condition $Y(i,0)=Y(i,n+1)=0$ translates into
$a_0=a_{n+1}=0$.  It is clear that these initial conditions together with
\eqref{xsys} determine all the $X(i,j)$'s. Moreover, the relations of
\eqref{xsys} impose no relations no the $a_i$'s. Therefore $a_1,\ldots,a_n$
can be considered as parameters of the solutions of the system
\eqref{xsys}.

By successively applying
\eqref{xsys} we can express the variables $X(i+2,i)$, $i=1,\dots, n$ in
terms of the $a_i$'s. The result is surprisingly simple:
\beq \label{solx}
X(i+2, i) = \frac{1-a_1 a_2\cdots a_i}{1-a_1 a_2\cdots a_{i+1}}
\end{equation}
To prove this formula, it is enough to check that relations \eqref{xsys}
hold identically if we substitute \eqref{solx} into them. To this end,
consider \eqref{xsys} for $i=j+1$ and substitute the above expression for
$X(j+1,j-1)$ and $X(j+2,j)$, and also $a_j$ and $a_{j+1}$ for $X(j,j)$ and
$X(j+1,j+1)$ respectively. After simple manipulations we obtain
$$\frac{1-a_j}{a_j}\cdot \frac{a_1a_2\cdots
a_j(1-a_{j+1})}{1-a_1a_2\cdots a_j} =
(a_{j+1}-1)\frac{a_1a_2\cdots a_{j-1}(a_j-1)}{1-a_1a_2\cdots a_j},$$
which is an obvious identity. This proves formula \eqref{solx}.

An important corollary of \eqref{solx} is that
\beq \label{diag}
X(n+2,n)=1-a_1a_2\cdots a_n,
\end{equation}
since $a_{n+1}=0$ by definition.

Now observe that we can apply the same recursion "starting at the other
end", i.e. do the calculations symmetrically with respect to the point
$((n+1)/2,(n+1)/2)$. Then we obtain an expression for $X(n-i-1,n-i+1)$,
which is equal to the expression for $X(i+2,i)$ with $a_i$ replaced
by $a_{n+1-i}$:
$$X(n-i-1,n-i+1) = \frac{1-a_n a_{n-1}\cdots a_{n+1-i}}{1-a_n a_{n-1}\cdots
a_{n-i}}.$$ In particular, when $i=n$ we have
\beq \label{diag1}
X(-1,1)=1-a_na_{n-1}\cdots a_1,
\end{equation} and thus $X(n+2,n)=X(-1,1)$ or $X(S(-1,1))=X(-1,1)$
(see the picture below in the case $n=4$).

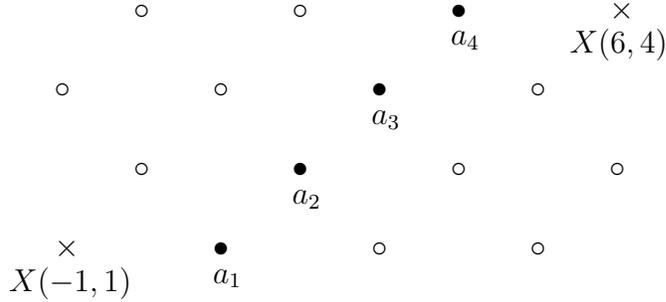
\begin{picture}(300,225)(0,-120)

\multiput(77,10)(30,30){2}{$\circ$}
\multiput(107,-20)(30,30){3}{$\circ$}
\put(77,-50){$\times$}
\put(60,-63){$X(-1,1)$}
\multiput(137,-50)(30,30){4}{$\bullet$}
\put(137,-60){$a_1$}
\put(167,-30){$a_2$}
\put(197,0){$a_3$}
\put(227,30){$a_4$}
\multiput(197,-50)(30,30){3}{$\circ$}
\put(287,40){$\times$}
\put(272,27){$X(6,4)$}
\multiput(257,-50)(30,30){2}{$\circ$}
\end{picture}

Note that the key fact is that formula \eqref{diag} for $X(n+2,n)$ is
invariant with respect to the involution $a_i \rightarrow a_{n+1-i}$.

Since the system \eqref{xsys} is invariant with respect
to simultaneous shifts of the coordinate $i$, we conclude that
$X(i,1)=X(S(i,1))$ for all $i\in\Z$, i.e. the periodicity holds for
$X(i,1)$, $i\in\Z$.
Clearly, these variables determine all the others via \eqref{xsys}. Now
note that \eqref{xsys} is invariant under $S$, thus the periodicity
holds for all of the $X(i,j)$'s and hence for all the $Y(i,j)$'s.\qed

\section{Properties of the dilogarithm functions.}    \label{property}
We consider two different types of dilogarithm functions: the Rogers
dilogarithm and the Bloch-Wigner dilogarithm. In this section, we discuss a
universal property of these functions, which will allow us to prove the
dilogarithm identities \eqref{gt} and their generalizations. In the case of
the Bloch-Wigner function, this property was first proved by S.~Bloch
\cite{Bloch}.

First let us introduce some notation.

For an abelian group $A$ consider the additive abelian group $A
\otimes_{\Z} A$. It consists of finite sums $$\sum_i n_i \cdot g_i \otimes
h_i, \quad \quad g_i, h_i \in A, n_i \in \Z,$$ with the obvious addition,
and subject to the relations $$(f g) \otimes h = f \otimes h + g
\otimes h, \quad \quad h \otimes (f g) = h \otimes f + h \otimes g,$$
$$1 \otimes h = h \otimes 1 = 0,$$ $$f^{-1} \otimes g = - f \otimes g,
\quad \quad g \otimes f^{-1} = - g \otimes f.$$

Denote by $S^2 A$ the subgroup of $A \otimes_{\Z} A$ generated by
elements of the form $a \otimes b + b \otimes a$ for all $a,b \in A$.

\subsection{The first identity for the Rogers dilogarithm.}
Let $C$ be the multiplicative group of nowhere vanishing continuous
differentiable functions from $[0,1]$ to $\R_+ = (0,+\infty)$.
Denote by  $L(z)$ the Rogers dilogarithm function defined on the interval
$[0,1]$ by formula \eqref{rogers}.

\begin{prop}    \label{id1}
Let $f_1,\ldots,f_N$ be continuous differentiable functions $[0,1] \arr
(0,1)$, such that $$\sum_{i=1}^N f_i \otimes (1-f_i) \in S^2 C.$$ Then
$$\sum_{i=1}^N L\left( f_i(x) \right) = \on{const}$$ as a function of $x
\in [0,1]$.
\end{prop}

\begin{pf}
By the definition of the Rogers dilogarithm function \eqref{rogers}
we have for $0 < z < 1$:
\begin{equation}    \label{differential}
d \, L(z) = \frac{1}{2} \left( \log z \cdot d \log (1-z) - \log (1-z) \cdot
d \log z \right).
\end{equation}
This implies that
\begin{multline}    \label{difsum}
d \sum_{i=1}^N L\left( f_i(x) \right) \\ = \frac{1}{2} \sum_{i=1}^N \left(
\log f_i(x) \cdot d \log (1-f_i(x)) - \log (1-f_i(x)) \cdot d \log
f_i(x) \right).
\end{multline}

If the $f_i$'s satisfy the condition of the proposition,
then there exist $g_j,h_j \in C, j=1,\ldots,m$, such that
\begin{equation}    \label{sym}
\sum_{i=1}^N f_i \otimes (1-f_i) = \sum_{j=1}^m \left( g_j \otimes h_j +
h_j \otimes g_j \right).
\end{equation}

For $x,y \in [0,1]$ we can define a homomorphism $\on{Log}_{x,y}: C \otimes
C \arr \R$, where $\R$ is considered as an additive group, by the formula
$\on{Log}_{x,y}(f \otimes g) = \log f(x) \cdot \log g(y)$. By applying the
homomorphism $\on{Log}_{x,y}$ to both sides of formula \eqref{sym}, we
obtain
$$\sum_{i=1}^N \log f_i(x) \log (1-f_i(y)) = \sum_{j=1}^m \log g_j(x) \log
h_j(y) + \log h_j(x) \log g_j(y).$$ Hence
\begin{multline}    \label{dif1}
\sum_{i=1}^N d \log f_i(x) \cdot \log (1-f_i(y)) \\ = \sum_{j=1}^m d \log
g_j(x) \cdot \log h_j(y) + d \log h_j(x) \cdot \log g_j(y),
\end{multline}
and
\begin{multline}    \label{dif2}
\sum_{i=1}^N \log f_i(x) \cdot d \log (1-f_i(y)) \\ = \sum_{j=1}^m \log
g_j(x) \cdot d \log h_j(y) + \log h_j(x) \cdot d \log g_j(y).
\end{multline}
Substracting \eqref{dif1} from \eqref{dif2} and setting $y=x$ we obtain
$$\sum_{i=1}^N \left( \log f_i(x) \cdot d \log (1-f_i(x)) - \log (1-f_i(x))
\cdot d \log f_i(x) \right) = 0,$$ which implies by \eqref{difsum} that
$$d \sum_{i=1}^N L\left( f_i(x) \right) = 0.$$
\end{pf}

\begin{cor}    \label{01}
Let $f_1,\ldots,f_N$ be as in \propref{id1}. Then $$\sum_{i=1}^N L(f_i(0))
= \sum_{i=1}^N L(f_i(1)).$$
\end{cor}

\subsection{The second identity for the Rogers function.}
The Rogers dilogarithm function can be extended to the whole real axis as
follows. Set
\begin{equation}    \label{pos}
L(z) = \frac{\pi^2}{3} - L\left( \frac{1}{z} \right), \quad
\quad z>1,
\end{equation}
and
\begin{equation}    \label{neg}
L(z) = - L\left( \frac{z}{z-1} \right), \quad \quad z<0.
\end{equation}
This is a continuous function on $\R$, which is differentiable for all $z$
except $0$ and $1$. It follows from this definition that $$\lim_{z \arr
+\infty} L(z) = \frac{\pi^2}{3}, \quad \quad \lim_{z \arr -\infty} =
-\frac{\pi^2}{6}.$$ These limits differ by $\pi^2/2$, and hence
$L(z)$ can be extended to $\infty$ as a function with values in
$\R/(\pi^2/2)\Z$. This way we obtain a function from $\R\Pro = \R
\cup \{ \infty \}$ to $R/(\pi^2/2)\Z$, which we denote by ${\cal
L}(z)$. The function ${\cal L}(z)$ is continuous and differentiable for all
$z$ except $0$ and $1$.

Now let ${\cal C}$ be the multiplicative group of non-zero rational
functions $[0,1] \arr \R$.

\begin{prop}    \label{id2}
Suppose that $f_1,\ldots,f_N \in {\cal C}$ are non-constant functions, such
that $$\sum_{i=1}^N f_i \otimes (1-f_i) \in S^2 {\cal C}.$$ Then
$$\sum_{i=1}^N {\cal L}(f_i(x)) = \on{const}$$ as a function of $x \in
[0,1]$.
\end{prop}

\begin{pf} The proof goes along the lines of the proof of
\propref{id1}. We consider each $f_i(x)$ as a continuous function $[0,1]
\arr \R\Pro$. For generic $x \in [0,1]$, the value $f_i(x)$ lies in one of
the intervals: $(0,1)$, $(1,+\infty)$, or $(-\infty,0)$. Therefore the
function $\sum_{i=1}^N {\cal L}(f_i(x))$ is differentiable for generic
$x$. We want to show that its differential vanishes.

Let us show that if $f_i(y)$ belongs to one of the intervals, then $$2 d \,
L(f_i(y)) = \log |f_i(y)| \cdot d \log |1-f_i(y)| - \log |1-f_i(y)| \cdot d
\log |f_i(y)|.$$

We consider the three cases separately. If $f_i(y) \in (0,1)$ then this
follows from formula \eqref{differential}.

If $f_i(y) \in (1,+\infty)$, we have by formulas \eqref{pos} and
\eqref{differential}: $$2 d L(f_i(y)) = - 2 d L\left( \frac{1}{f_i(y)}
\right) $$ $$= \log \left( \frac{1}{f_i(y)} \right) \cdot d \log \left(
1-\frac{1}{f_i(y)} \right) - \log \left( 1-\frac{1}{f_i(y)} \right) \cdot d
\log \left( \frac{1}{f_i(y)} \right)$$ $$= \log |f_i(y)| \cdot d \log
|1-f_i(y)| - \log |1-f_i(y)| \cdot d \log |f_i(y)|.$$

If $f_i(y) \in (-\infty,0)$, we have by formulas \eqref{neg} and
\eqref{differential}: $$2 d L(f_i(y)) = - 2 d L\left(
\frac{f_i(y)}{f_i(y)-1} \right)$$
$$= \log \left( \frac{f_i(y)}{f_i(y)-1} \right) \cdot d
\log \left( 1-\frac{f_i(y)}{f_i(y)-1} \right) - \log \left(
1-\frac{f_i(y)}{f_i(y)-1} \right) \cdot d \log \left(
\frac{f_i(y)}{f_i(y)-1} \right)$$ $$= \log
|f_i(y)| \cdot d \log |1-f_i(y)| - \log |1-f_i(y)| \cdot d \log |f_i(y)|.$$

Now for $x \in [0,1]$ let ${\cal C}_x$ be the subgroup of ${\cal C}$ which
consists of those functions, which have neither zero or pole at $x$. Define
for $x,y \in [0,1]$ a homomorphism ${\cal Log}_{x,y}: {\cal C}_x
\otimes {\cal C}_y \arr \R$ by the formula ${\cal Log}_{x,y}(f
\otimes g) = \log |f(x)| \cdot \log |g(y)|$. Using this homomorphism and
the formulas above in the same way as in the proof of \propref{id1}, we
obtain that $d \sum_{i=1}^N L(f_i(x)) = 0$ for generic $x \in
[0,1]$. Therefore by continuity, $\sum_{i=1}^N {\cal L}(f_i(x)) =
\on{const}$ for all $x \in [0,1]$. \end{pf}

\begin{cor}    \label{01mod}
Let $f_1,\ldots,f_N$ be as in \propref{id2}. Then
$$\sum_{i=1}^N L(f_i(0)) = \sum_{i=1}^N L(f_i(1)) \; \mod
\frac{\pi^2}{2}.$$
\end{cor}

\subsection{Identity for the Bloch-Wigner function.}
The Bloch-Wigner function is the function $D: \C \arr \R$ given by the
formula
\begin{equation}    \label{bw}
D(z) = -\on{Im} \int_0^z \log (1-w)\,d\log w + \log|z| \on{Arg} (1-z).
\end{equation}
This function is single-valued and real analytic everywhere
except for $z=0,1$, where it is only continuous \cite{Bloch}.

Let $\goth{C}$ be the multiplicative group of non-vanishing holomorphic
functions ${\cal D}_R \arr \C$, where ${\cal D}_R$ is the disc of radius
$R>1$.

\begin{prop}    \label{idbw}
Suppose that $f_1,\ldots,f_N$ are holomorphic functions such that $f_i,
1-f_i \in \goth{C}$ for all $i=1,\ldots,N$, and $$\sum_{i=1}^N f_i \otimes
(1-f_i) \in S^2 \goth{C}.$$ Then
$$\sum_{i=1}^N D(f_i(x)) = \on{const}$$ as a function of $x \in {\cal
D}_R$.
\end{prop}

The proof is similar to the proofs of the previous identities. Let us
observe that $$d D(z) = \log|z|\, d \on{Arg}(1-z) - \log|1-z|\, d\on{Arg}
z.$$ To prove the proposition we should consider the homomorphism
$\goth{Log}_{x,y}: \goth{C} \otimes \goth{C} \arr \R \otimes (\R/2\pi)$
given by $\goth{Log}_{x,y}(f \otimes g) = \log|f(x)| \otimes \on{Arg} g(y)$
and proceed in the same way as above (see also \cite{Bloch}).

\begin{cor}[\cite{Bloch}]    \label{01bw}
Let $f_1,\ldots,f_N$ be as in \propref{idbw}. Then $$\sum_{i=1}^N D(f_i(0))
= \sum_{i=1}^N D(f_i(1)).$$
\end{cor}

\section{The dilogarithm identities.}
In this section we prove the following result.

\begin{thm}    \label{osnovnaya}
\begin{enumerate}
\item[(1)]
Suppose that the real numbers $X(i,j), i\in\Z, j=\onen$,
satisfy the system of equations \eqref{xsys}. Then if all $X(i,j) \in
(0,1)$,
\begin{equation}    \label{dilid1}
\sum_{i=1}^{n+3} \sum_{j=1}^{n} L(X(i,j)) = \frac{\pi^2}{6} n(n+1),
\end{equation}
and in general
\begin{equation}    \label{dilid2}
\sum_{i=1}^{n+3} \sum_{j=1}^{n} L(X(i,j)) = \frac{\pi^2}{6} n(n+1) \quad
\mod \frac{\pi^2}{2}.
\end{equation}

\item[(2)]
Suppose that the complex numbers $X(i,j), j\in\Z, i=\onen$,
satisfy the system of equations \eqref{xsys}. Then
\begin{equation}    \label{dilid3}
\sum_{i=1}^{n+3} \sum_{j=1}^{n} D(X(i,j)) = 0.
\end{equation}
\end{enumerate}
\end{thm}

First we prove a generalization of \eqref{diag}. For $\ep=0,1$ let
$$\goth{S}_\ep =\{(i,j)|\,i \in \Z, 1\leq j\leq n,\, i+j=\ep\mod 2\}.$$ Let
$\ca_\ep$ be the coset $\goth{S}_\ep/\{ S(i,j)\sim (i,j) \},$ where $S$ is
the transformation introduced in \secref{per}, and $p_\ep: \goth{S}_\ep
\arr \ca_\ep$ be the corresponding projection. According to \thmref{peri},
if the $X(i,j)$'s satisfy the system \eqref{xsys}, then $X(i,j) =
X(S(i,j))$. Therefore the $X(i,j)$'s can be considered a function on $\ca_0
\cup \ca_1$.

For $(i,j) \in \goth{S}_\ep$, introduce the cone
$\Cone_\ep(i,j) \subset \ca_\ep$ as the image of the set
$$\{(i',j')\in\goth{S}_\ep|\,|i'-i|\geq |j'-j|\}$$ under the map
$p_\ep$. We note that the restriction of $p_\ep$ to $\Cone_\ep(i,j)$ is
injective. Finally, denote by
$\bcone_\ep(i,j)=\ca_\ep\backslash\Cone_\ep(i,j)$ the cone's complement.
\begin{prop}     \label{cone}
For any $(i,j) \in \ca_\ep$,
\beq \label{total}
1-X(i,j) = \prod_{(i',j')\in\bcone_\ep(i,j)}X(i',j')
\end{equation}
\end{prop}
{\it Proof.} We prove the proposition by induction on $j$. For $j=1$, the
set $\bcone_\ep(i,1)$ is the image of the set $\{ (i+l+1,l)\in\goth{S}_\ep
\}$ under the map $p_\ep$, and \eqref{total} is equivalent to \eqref{diag1}.

Assume now that the proposition is true for all $j<k$. For a fixed $i$ the
relation \eqref{xsys} between $X(i-1,k-1),X(i+1,k-1),X(i,k-2)$ and $X(i,k)$
gives
\beq \label{xsysk} 1-X(i,k) = \frac{1}{1-X(i,k-2)} \cdot
\frac{X(i-1,k-1)}{1-X(i-1,k-1)} \cdot \frac{X(i+1,k-1)}{1-X(i+1,k-1)}.
\end{equation}
By our inductive assumption, we can make the substitution \eqref{total} for
the factors of the form $1-X$ in the right hand side. Then we see that the
validity of formula \eqref{total} for $X(i,k)$ is equivalent to the
statement
$$ \bcone(i-1,k-1)+\bcone(i+1,k-1) =\bcone(i,k-2)+\bcone(i,k)+
\{(i-1,k-1),(i+1,k-1)\},$$ where by addition of sets we mean the union of
their elements counted with multiplicities. This last statement is a simple
fact of elementary geometry.

Note that at the two ends, when $k=1$ or $n+1$, one of the factors in
\eqref{xsysk} is missing. This corresponds to the fact that
$\Cone(i,0)=\Cone(i,n+1)=\ca_n$ and hence $\bcone_\ep(i,0) =
\bcone_\ep(i,n+1) = \emptyset.$  \qed

\bigskip
\noindent{\em Proof of \thmref{osnovnaya}.} We show that any solution
of \eqref{xsys} can be connected to a particular solution $X(i,j)_0$ for
which the the sum of values of the dilogarithm functions is known.

The solution $X(i,j)_0$ corresponds to the UV limit, i.e. $X(i,j)_0$ is
$i$--independent. Therefore the system \eqref{xsys} becomes:
$$X(i,j)^2 = \prod_{l=1}^n (1-X(i,j))^{A_{jl}},$$ where $(A_{ij})$ is the
Cartan matrix of type $A_n$. The following formula gives a particular
solution of this system:
\begin{equation}    \label{basic}
X(i,j)_0 = 1 - \frac{\sin^2 \frac{\pi}{n+3}}{\sin^2
\frac{\pi(i+1)}{n+3}}, \quad i\in\Z; j=\onen
\end{equation}
(see e.g. \cite{FS}).

Solutions of the system \eqref{xsys} are in one-to-one correspondence with
the numbers $X(0,j)$ and $X(1,j)$, where $j=\onen$. We can choose them
arbitrarily, and if none of them is equal to $0$ or $1$, then all other
numbers $X(i,j)$ can be uniquely determined recursively using the system
\eqref{xsys}.

{}From now on we restrict our attention to the identity \eqref{dilid1}. The
identities \eqref{dilid2} and \eqref{dilid3} can be treated in the same
fashion.

We have: $X(i,j)_0 \in (0,1)$. Let $\{ X(i,j)_1 \}$ be another solution of
\eqref{xsys}, such that all $X(i,j)_1 \in (0,1)$. Consider the functions
$$X(i,j)(z) = zX(i,j)_1 + (1-z)X(i,j)_0, \quad \quad i=0,1;\quad j=\onen$$
Using these functions, we can determine uniquely all other functions
$X(i,j)(z)$ recursively from the system \eqref{xsys}. The functions
$X(i,j)(z)$ constructed this way satisfy the system \eqref{xsys} for all $z
\in [0,1]$ and $X(i,j)(0) = X(i,j)_0, X(i,j)(1) = X(i,j)_1$ for all $i,j$.

\begin{lem}    \label{between}
Let $X(i,j)=X(i,j)(z), i\in\Z, j=\onen$, be functions $[0,1] \arr \R$,
which satisfy the system \eqref{xsys} for all $z \in [0,1]$. Suppose that
\begin{enumerate}
\item[(1)] $0 < X(i,j)(0) < 1$ for all $i,j$;

\item[(2)] $X(i,j)(z), i=0,1$, are continuous differentiable functions;

\item[(3)] $0 < X(i,j)(z) < 1$ for $i=0,1$ and all $z \in [0,1]$.
\end{enumerate}

Then $X(i,j)(z)$ are continuous differentiable functions such that $0 <
X(i,j)(z) < 1$ for all $i,j$ and $z \in [0,1]$.
\end{lem}

\begin{pf} We prove the lemma by induction on $i$. We already know
that it holds for $i=0,1$. Suppose that it holds for $0\leq i <l$.  Using
the system \eqref{xsys} we can express $X(l,j)$ via $X(l-2,j)$,
$X(l-1,j+1)$ and $X(l-1,j-1)$ as follows:
\begin{equation}    \label{xsysl}
1-X(l,j)^{-1} = \frac{(1-X(l-1,j+1))(1-X(l-1,j-1))X(l-2,j)}{X(l-2,j)-1}.
\end{equation}
We know that $X(l,j)(0) = X(i,j)_0 \in (0,1)$, and it is clear from
\eqref{xsysl} that $X(l,j)$ can not be equal to $0$ or $1$ if all
$X(l-2,j)$, $X(l-1,j+1)$ and $X(l-1,j-1)$ lie in the interval
$(0,1)$. Therefore by our inductive assumption we have: $X(l,j)(z) \in
(0,1)$ for all $z \in [0,1]$ and $X(l,j)(z)$ is continuous and
differentiable. Hence the lemma holds for all $i\geq 0$. The case $i<0$ is
treated similarly.
\end{pf}

According to this lemma, the functions $X(i,j)=X(i,j)(z)$ that we have
constructed, belong to the group $C$. We also know that the values of
$X(i,j)(z)$ at $0$ and $1$ are equal to $X(i,j)_0$ and $X(i,j)_1$,
respectively.

Note that there is a one-to-one correspondence between the coset $\ca_\ep$
and the set $\{ (i,j) \in \goth{S}_\ep|\, 1\leq i\leq n+3, 1\leq j\leq n
\}$. Hence
\begin{multline}    \label{sym1}
\sum_{i=1}^{n+3} \sum_{j=1}^n X(i,j) \otimes (1-X(i,j)) \\ = \sum_{(i,j)
\in \ca_0} X(i,j) \otimes (1-X(i,j)) + \sum_{(i,j) \in \ca_1} X(i,j)
\otimes (1-X(i,j)).
\end{multline}
The functions $X(i,j)=X(i,j)(z)$ satisfy the system \eqref{xsys}
for all $z \in [0,1]$. By expressing $1-X(i,j)$ in terms of $X(i',j')$'s
using formula \eqref{total} we obtain:
\begin{equation}    \label{sym2}
\sum_{(i,j) \in \ca_\ep} X(i,j) \otimes (1-X(i,j)) = \sum_{(i,j) \in
\ca_\ep} \; \sum_{(i',j') \in \bcone_\ep(i,j)} X(i,j) \otimes X(i'j').
\end{equation}

Since  the relation
$$\{((i,j),(i',j'))|\,
(i',j')\in\bcone_\ep(i,j)\}\subset\ca_\ep\times\ca_\ep$$ is symmetric, we
conclude from formulas \eqref{sym1} and \eqref{sym2} that
$$\sum_{i=1}^{n+3} \sum_{j=1}^n X(i,j) \otimes (1-X(i,j)) \in S^2 C.$$

Therefore these functions satisfy the conditions of \corref{01}. Hence the
sum of the values of the Rogers dilogarithm function at $X(i,j)_0$ equals
to that at $X(i,j)_1$. But we can derive from the known dilogarithm
identity \cite{Lewin,proof,FS}
$$\sum_{j=1}^n L\left( \frac{\sin^2 \frac{\pi}{n+3}}{\sin^2
\frac{(j+1)\pi}{n+3}} \right) = \frac{\pi^2}{6} \frac{2n}{n+3}$$ and the
Euler identity $L(z) + L(1-z) = \pi^2/6$ that
$$\sum_{i=1}^{n+3} \sum_{j=1}^{n} L(X(i,j)_0) = \frac{\pi^2}{6} n(n+1).$$
The identity \eqref{dilid1} now follows from \corref{01}.

By using the same argument and \corref{01mod} we obtain a proof of
\eqref{dilid2}.

Finally, observe that $D(z)=0$ for all real $z$, and hence
$$\sum_{i=1}^{n+3} \sum_{j=1}^{n} D(X(i,j)_0) = 0.$$ The identity
\eqref{dilid3} now follows from \corref{01bw}.\qed

\end{document}